\documentclass[11pt]{article}
\usepackage{amsmath,amssymb,epsfig, latexsym}
\begin{document}

\newcommand{\beg}{\begin{equation}\label}
\newcommand{\enq}{\end{equation}}
\newcommand{\p}{\partial}
\newcommand{\pr}{\prime}
\newcommand{\bib}{\bibitem}
\newcommand{\nabl}{{\bf \nabla}}

\begin{center}
{\bf NATURE OF TIME AND PARTICLES-CAUSTICS:} \\

{\bf PHYSICAL WORLD IN ALGEBRODYNAMICS AND IN TWISTOR THEORY}

\vskip 1cm

Vladimir~V.~Kassandrov\vskip 1 cm

\end{center}
{\footnotesize Department of General Physics, People's Friendship University of 
Russia,Ordjonikidze 3, 117419 Moscow, Russia\\ 
E-mail: vkassan@sci.pfu.edu.ru}\vskip 1cm

\noindent
{\footnotesize {\bf Abstract.}

\noindent
In the field theories with twistor structure particles can be identified with 
(spacially bounded) caustics of null geodesic congruences defined by the twistor 
field. As a realization, we consider the ``algebrodynamical'' approach based
on the field equations which originate from noncommutative analysis (over the 
algebra of biquaternions) and lead to the complex eikonal field and to the set 
of gauge fields associated with solutions of the eikonal equation. Particle-like 
formations represented by singularities of these fields possess ``elementary''
electric charge and other realistic ``quantum numbers'' and manifest 
self-consistent time evolution including transmutations. 
Related concepts of generating ``World Function'' and of 
multivalued physical fields are discussed. 
The picture of Lorentz invariant light-formed aether and of matter 
born from light arises then quite naturally. The notion of the Time Flow 
identified with the flow of primodial light (``pre-Light'') is introduced in 
the context.}

\section{Introduction. The algebrodynamical field theory}
\label{sec1}

Theoretical physics has arrived to the crucial point at which it should 
fully reexamine the sense and the interrelations of the three fundamental 
entities: fields, particles and space-time geometry. 
{\it String theory} offers a way to derive the low-energy phenomenology from 
the unique physics at Plankian scale. However, it doesn't claim to find the 
{\it origin} of physical laws, the {\it Code of Universe} and is in fact nothing 
but one more attempt to {\it describe} Nature (in a possibly the most effective 
way) but not at all to {\it understand} it.  

{\it Twistor program} of R. Penrose~\cite{twistpen,penrose} suggests 
an alternative to string theory in the framework of which one can hope, 
in principle, to explain the origin of basic physical entities. 
For this, one only assumes the existence of 
the primary {\it twistor space} $\mathbb{C}P^3$ which underlies the physical 
space-time and predetermines its Minkowsky geometry and, to some extent, 
the set of physical fields. 

The most interesting manifestation 
of twistor structure is its ability to reduce the resolution of free 
massless (conformally invariant) equations (both linear and nonlinear ones,  
specifically of the Yang-Mills type) either to explicit integration in twistor 
space (the so called Penrose transform) or to resolution of purely 
algebraic problems (the Kerr theorem, the Ward construction etc.~\cite{penrose}). 
Making use of the Kerr theorem and of the Penrose's ``nonlinear graviton 
construction'', one can also obtain, in a purely algebraic way, the whole set 
of the self-dual solutions to (complex) Einstein equations.

However, general concept of twistor program as a unified field 
theory is not at all clear or formulated up to now. Which equations are 
really fundamental, in which way can the massive fields  be described  
and in which way the particles' spectrum can be obtained? And, finally, 
why precisely twistor, a rather refined mathematical object, should be taken 
as a basis of fundamental physics? 

In the interim, twistor structure arises quite naturally in the so called 
{\it algebrodynamics} of physical fields which has been developed in our 
works. From general viewpoint, the paradigm of algebrodynamics 
can be thought of as a revive of Pithagorean or 
Platonean ideas about {\it ``Numbers governing physical laws''}. As the 
only (!) postulate of algebrodynamics one admits the existence of a certain 
unique and exeptional structure, of purely abstract (algebraic) nature, the 
internal properties of which completely determine both the 
geometry of physical space-time and the dynamics of physical fields 
(the latters being also algebraic in nature). 

In the most successful realization of algebrodynamics principal structure of the 
``World algebra'' has been introduced via generalization of complex analysis to 
exeptional noncommutative algebras of quaternion ($\mathbb{Q}$) type~\cite
{AD,GR,acta,prot,jos}.  
In particular, it was demonstrated that explicit account of noncommutativity in 
the very definition of functions ``differentiable'' in $\mathbb{Q}$ 
inevitably results in the {\it non-linearity} of the generalized Cauchy-Riemann 
equations (GCRE) which follow. This makes it possible to regard the GCRE as 
fundamental dynamical equations of {\it interacting} physical fields 
represented by (differentiable) functions of the algebraic $\mathbb{Q}$-type 
variable. 

A wide class of such fields-functions exists only for the complex extension of 
$\mathbb{Q}$-algebra, i.e. for the algebra of complex quaternions $\mathbb{B}$ 
({\it biquaternions}). Over the $\mathbb{B}$-algebra, the GCRE turn to be 
Lorentz invariant and acquire, moreover, the gauge and the spinor 
structures. On this base a self-consistent and unified {\it algebrodynamical 
field theory} has been constructed in our works~\cite{AD,GR,kasan,prot,jos,trish,
vest,eik}. 

From the physical viewpoint, the most important property of GCRE is their 
direct correspondence to a fundamental  {\it light-like} structure. 
The latter manifests itself in the fact that every (spinor) component 
$S(x,y,z,t) \in \mathbb{C}$ of the primary $\mathbb{B}$-field must satisfy the 
{\it complex eikonal equation} (CEE) ~\cite{wilson,AD}
\beg{eik}
\eta^{\mu\nu}\p_\mu S \p_\nu S = 
(\p_t S)^2 - (\p_x S)^2 - (\p_y S)^2 - (\p_z S)^2 = 0 ,
\enq
where $\eta_{\mu\nu}=diag\{1,-1,-1,-1\}$ is the Minkowsky metric and $\p$ 
stands for the partial derivative by respective coordinate. The CEE (\ref{eik}) 
is Lorentz invariant, nonlinear and plays the role similar to that 
of the {\it Laplace equation} in complex analysis. Each solution to GCRE 
can be reconstructed from a set of (four or less) solutions to CEE. 

In the meantime, in~\cite{eik} the intrinsic {\it twistor} structure of CEE 
has been discovered, and on its base the general solution of the nonlinear 
eikonal equation  
has been obtained. It was proved that, in this respect, every CEE solution 
belongs to one of two classes which both can be obtained from a twistor generating 
function via a simple and purely algebraic procedure. This construction allows
also for definition of singular loci of the null geodesic congruences correspondent 
to the eikonal field -- the {\it caustics}. Just at the caustics -- the envelopes 
of the congruences -- the neighbouring rays intersect each other, and the 
associated physical fields turn to infinity forming, thus, a unique {\it 
particle-like} object -- a common source of the fields and of the congruence itself. 
Thus, in the algebrodynamical theory {\it the particles can be 
considered as (spacially bounded) caustics of the primodial null congruences}.

On the other hand, null congruences naturally define the universal local 
``transfer'' of the basic twistor field with fundamental constant velocity 
``c'' (in full analogy with the transfer of field by an electromagnetic wave)
and point thus to exceptional role of the time coordinate in the algebrodynamical 
scheme and in twistor theory in general. 
Existence of the ``Flow of Time'' becomes therein a direct consequence of the 
existence of Lorentz invariant 	``aether'' formed by the primodial
light-like congruence (``preLight''). In the paper, we underline the principal 
property of {\it multivaluedness} of the fundamental complex solution to CEE 
(``World solution'') and of the physical fields associated with it. As a result, 
at each space-time point one has a {\it superposition} of a great number of
rays which belong to locally distinct null congruences, and the Time Flow turns 
to be {\it multi-directional}, i.e. consists of a number of superposed 
``subflows'' (linked globally by complex structure into a unique physical 
``corpuscular-field'' dualistic complex). 

In section 2 we consider the twistor structure of CEE and the procedure of 
algebraic construction of its two classes of solutions. A few simple 
illustrative examples are presented. In section 3 we
discuss the caustic structure of the CEE solutions, in particular of spatially 
bounded type (particle-like singular objects), and the properties of associated 
physical fields. In section 4, we introduce the ``World function'' responsible 
for generation of the ``World solution'' to CEE and discuss the related
concept of multivaluedness of physical fields. Final section 5 is devoted to 
some general issues which bear on the nature of physical time. The notions of 
the {\it primodial light} (``pre-Light'') and of the {\it light-formed aether} 
are introduced, and the Time Flow is actually identified with the Flow of
preLight. Intrinsic structure of these fundamental flows is studied which  
relates to the property of multivaluedness of the basic twistor field.

The article is a continuation of our paper~\cite{number}. In order to simplify 
the presentation, we avoid to apply the 2-spinor and the other refined 
mathematical formalisms, for this refering a prepared reader
to our recent papers~\cite{prot,jos,vest,eik}.
  
\section{The two classes of solutions to the complex eikonal equation}

The eikonal equation describes the process of propagation of wave fronts (field 
discontinuties) in any relativistic theory, in Maxwell electrodynamics in 
particular~\cite{fock,petrov}. Physical and mathematical problems related to the 
eikonal equation were dealt with in a lot of works, see e.g.~\cite{bate,newman,
vinogr,arnold,arnold2,kruzhkov}. 

The complex eikonal equation (CEE) arises naturally in 
problems of propagation of restricted light beams~\cite{maslov} and in theory of 
congruences related to solutions of Einstein or Einstein-Maxwell system of 
equations~\cite{wilson}. We, however, interpret the complex eikonal, 
to the first turn, as a fundamental physical field 
which describes, in particular, the interacting and 
``self-quantized'' particle-like objects formed by singularities of the CEE
solutions. By this, the electromagnetic and the other conventional physical 
fields can be associated with any solution of the CEE; they are responsible 
for description of the process of interaction of particles-singularities.
Note that particle-like properties of field singularities related to the 
{\it 5-dimensional} real eikonal field have been studied in ~\cite{pyt'iev}; 
the concept of particles as singularities of electromagnetic and eikonal fields has been 
incidentically discussed by many authors, in particular by H. Bateman
~\cite{bate} af far as in 1915. 

We start with a definition, together with Cartesian space-time coordinates 
$\{t,x,y,z\}$, of the so called {\it spinor} or {\it null} coordinates 
$\{u,v,w,\bar w\}$ (the light velocity is taken to be unity, $c=1$)
\beg{spcoord}
u = t+z, ~~v=t-z, ~~w=x-iy, ~~\bar w =x+iy 
\enq
which form the {\it Hermitian} $2\times 2$ matrix $X=X^+$ of coordinates
\beg{hermit}
X = \left(
\begin{array}{cc}                                                      
u & w \\  \bar w & v     
\end{array}
\right)
\enq
In the spinor coordinates representation the CEE (\ref{eik}) looks as follows:
\beg{eiksp}
\p_u S \p_v S - \p_w S \p_{\bar w} S = 0 .
\enq 

The CEE possesses a remarkable {\it functional invariance}~\cite{AD,GR}: for  
every $S(X)$ being its solution any (differentiable) function $f(S(X))$ is 
also a solution. The eikonal equation is known also~\cite{bate} to be invariant 
under transformations of the full 15-parameter {\it conformal group} of the 
Minkowsky space-time.

Let us take now an {\it arbitrary homogeneous} function $\Pi$ of two pairs of 
complex variables $\{\xi,\tau\}$ 
\beg{twist}
\Pi = \Pi(\xi_0, \xi_1, \tau^0, \tau^1) 
\enq 
which are {\it linearly dependent} at any space-time point via the so called 
{\it incidence relation}
\beg{inc}
\tau = X\xi  ~~\Leftrightarrow ~~\tau^0=u\xi_0 + w\xi_1, ~~\tau^1=\bar w\xi_0
+v\xi_1,
\enq
and which transform as {\it 2-spinors} under Lorentz rotations~\footnote{ 
To simplify the notation, we do not distinguish the primed and unprimed 
spinor indices. In the incidence relation (\ref{inc}) the standard factor ``i'' 
(imaginary unit) is omitted what is admissible under the proper redefinition of 
the twistor norm}.
The pair of 2-spinors $\{\xi(X),\tau(X)\}$ linked through Eq.(\ref{inc}) is known 
as a (null projective) {\it twistor} of the Minkowsky space-time~\cite{penrose}. 

Let us assume now that one of the components of the spinor $\xi(X)$, say $\xi_0$, 
is not zero. Then, by virtue of homogeneity of the function $\Pi$,  
we can reduce the number of its arguments to {\it three projective twistor 
variables}, namely to
\beg{proj}
\Pi = \Pi(G, \tau^0, \tau^1), ~~~~G=\xi_1 / \xi_0,~~\tau^0=u+wG,~~\tau^1=\bar w +vG 
\enq
Now we are in order to formulate the main result proved in our paper~\cite{eik}.

\noindent
{\bf Theorem.} {\it Any (analytical) solution of CEE belongs, with respect to its 
twistor structure, to one of two and only two classes and can be obtained from some 
generating twistor function of the form (\ref{proj}) via one of two simple 
algebraical procedures (described below)}

\noindent
(note only that for solutions with zero spinor component, $\xi_0=0$, another 
gauge, in compare with the one used above, should be choosed).
                                                                
To obtain the first class of solutions, let us simply resolve the algebraic 
equation defined by the function (\ref{proj}) 
\beg{kerr}
\Pi(G,u+wG, \bar w+vG) =0
\enq
with respect to the only unknown $G$. In this way we come to a complex field 
$G(X)$ which necessarily satisfies the CEE. Indeed, after substitution $G=G(X)$ 
Eq.(\ref{kerr}) becomes an {\it identity} and, in particular, can be differentiated 
with respect to the spinor coordinates $u,v,w,\bar w$. Then we get
\beg{deriv}
P \p_u G = -\Pi_0, ~~P\p_w G = -G\Pi_0, ~~P\p_{\bar w} G = -\Pi_1, ~~
P\p_v G = -G\Pi_1,
\enq
where $\Pi_0,\Pi_1$ are the partial derivatives of $\Pi$ with respect to its 
twistor arguments $\tau^0,\tau^1$ while $P$ is its {\it total} derivative 
with respect to $G$,
\beg{totder}
P = \frac{d\Pi}{dG} = \p_G \Pi + w\Pi_0 +v\Pi_1 ~,
\enq
which we thus far assume to be nonzero in the space-time domain considered. 
Multiplying then Eqs.(\ref{deriv}) we prove that $G(X)$ satisfies the 
CEE in the form (\ref{eiksp}). It is easy to check that                                                       
{\it arbitrary} twistor function $S=S(G,u+wG,\bar w+vG)$, under substitution of 
the obtained $G=G(X)$, also satisfies the CEE (owing to the functional constraint  
(\ref{kerr}) it depends in fact on only {\it two} of three twistor variables). 

To obtain the second class of CEE solutions, we have from the very beginning 
to differentiate the function $\Pi$ with respect to $G$ and only after this 
to resolve the resulting algebraic equation 
\beg{caust}
P=\frac{d\Pi}{dG} = 0 
\enq
with respect to $G$ again. Now the function $G(X)$ does not satisfy the CEE; 
however, if we substitute it into (\ref{proj}) the quantity $\Pi$ 
becomes an explicit function of space-time coordinates and necessarily 
satisfies the CEE (as well as any function $f(\Pi(X))$ by virtue 
of functional invariance of the CEE). Indeed, differentiating the function $\Pi$ 
with respect to the spinor coordinates we get
\beg{der2}
\p_u \Pi = \Pi_0 +P\p_u G, ~~\p_w \Pi = G\Pi_0+P\p_w G, ~~\p_{\bar w} \Pi =
\Pi_1+P\p_{\bar w} G, ~~\p_v \Pi = G\Pi_1 + P\p_v G, 
\enq 
and, taking into account the generating condition (\ref{caust}), we immegiately  
find that the function $\Pi$ itself obeys the CEE (\ref{eiksp}).

The functional condition (\ref{kerr}) and, therefore, the CEE solutions of the 
first class are in fact well known. Indeed, apart from the CEE, the field 
$G(X)$,  if it is obtained by the resolution of Eq.(\ref{kerr}), satisfies (as it is 
easily seen from Eqs.(\ref{deriv}) for derivatives), the {\it over-determined} 
system of differential constraints 
\beg{sfc}
\p_u G = G\p_w G, ~~~\p_{\bar w} G = G\p_v G
\enq
which define the so called {\it shear-free (null geodesic) congruences} (SFC). 
By this, algebraic Eq.(\ref{kerr}) represents (in implicit form) 
{\it general solution} of Eqs.(\ref{sfc}), i.e. describes the whole set of SFC in 
the Minkowski space-time. This remarkable statement proved in~\cite{kerr} is 
known as the {\it Kerr theorem}. 

The second class of CEE solutions generated by algebraic 
constraint (\ref{caust}), to our knowledge, hasn't been considered in 
literature previously
~\footnote{Study of solutions of the {\it real} eikonal equation by differentiation 
of generating functions depending on coordinates as parameters is used in 
general theory of sungularities of caustics and wavefronts~\cite{arnold2}}.                                             
It is known, however, that condition (\ref{caust}) defines the 
{\it singular locus} for SFC, i.e. for the CEE solutions obtained from the Kerr 
constraint (\ref{kerr}). Precisely, condition (\ref{caust}) fixes the 
{\it branching points} of the principal complex field $G(X)$ or, equivalently, 
-- the space-time points where Eq.(\ref{kerr}) has {\it multiple roots}. 
As to the CEE solutions of the second class themselves, their branching points 
occure at the locus defined by another condition which evidently follows from 
generating Eq.(\ref{caust}) and has the form
\beg{supsing}
\Lambda=\frac{d^2 \Pi}{dG^2} = 0.
\enq 
The null congruences (especially the congruences with zero shear), as well as 
their singularities and branching points, play crucial role in the 
algebrodynamical approach. They will be discussed below in more details. Here 
we only repeat that, as it has been proved in~\cite{eik}, 

\noindent
{\it the two simple generating procedures described above exhaust all the 
(analytical) solutions to the CEE representing, thus, its {\bf general solution}.}

The obtained 
result can be thought of as {\it a direct generalization of the Kerr theorem.} 
Below, in order 
to make the exposition more clear, we present several examples of the 
described construction.   

\noindent
{\bf 1. Static solutions}. Let the generating function $\Pi$ depends on 
its twistor variables in the following way:
\beg{stat}
\Pi=\Pi(G,H), ~~~H=G\tau^0-\tau^1=wG^2+2zG-\bar w,
\enq
where $z=(u-v)/2$, and the time coordinate $t=(u+v)/2$ is, in this way, 
eliminated. It is evident that the generating ansatz (\ref{stat}) covers the 
whole class of {\it static} CEE solutions.

In ~\cite{burin,wilson} it was proved that static solutions to the SFC equations 
(and, therefore, static solutions to the CEE too) with {\it spacially bounded} 
singular locus are exhausted, up to 3D translations and rotations, by the 
{\it Kerr solution}~\cite{kerr} which follows from generating function of the 
form
\beg{kerrgen}
\Pi=H+2iaG=wG^2+2z^* G-\bar w,  ~~~(z^*=z+ia)
\enq
with a real constant parameter $a\in \mathbb{R}$. Explicitly resolving equation 
$\Pi=0$ which is quadratic in $G$ we obtain the two ``modes'' of the field $G(X)$ 
\beg{kerrsol}
G=\frac{\bar w}{z^*\pm r^*}=\frac{x+iy}{z+ia \pm \sqrt{x^2+y^2+(z+ia)^2}}
\enq
which in the case $a=0$ correspond to the ordinary {\it stereographic 
projection} $S^2\mapsto \mathbb{C}$  from the North or the South pole 
respectively. It is easy to check that this solution and also its twistor 
counterpartners 
\beg{count}
\tau^0= t + r^* , ~~~\tau^1=G\tau^0 ,
\enq
satisfy the CEE (as well as any function of them). Correspondent SFC is 
in the case $a=0$ {\it radial} with a point singularity; in general case $a\ne 0$ 
the SFC is formed by the rectilinear constituents of a system of hyperboloids 
and has a {\it ring-like} singularity of a radius $R=\vert a \vert$. Using this 
SFC, a Riemannian metric (of the ``Kerr-Schild type'') and an electric field 
can be defined which satisfy together the electrovacuum Einstein-Maxwell system. 
In the case $a=0$ this is the Reissner-Nordstr\"om solution with Coulomb 
electric field, in general case -- the Kerr-Newman solution with three 
characteristical parameters: the mass $M$, the electric charge $Q$ and the 
angular momentum (spin) $Mca$, -- for which the field distribution possesses also 
the proper magnetic moment $Qa$ which corresponds to the gyromagnetic ratio 
specific for the Dirac particle~\cite{carter,lopes}. In the algebrodynamical 
scheme, moreover, {\it electric charge of the 
point or the ring singularity is necessarily fixed in modulus}, i.e. ``elementary'' 
~\cite{AD,GR,vest,sing} (see also~\cite{prot} where a detailed discussion 
of this solution in the framework of algebrodynamics can be found). 

Now let us obtain, from the same generating function, a solution to CEE of the 
second class. Differentiating Eq.(\ref{kerrgen}) with respect to $G$ and 
equating derivarive to zero, we get $G=-z^*/w$ and, substituting this 
expression into Eq.(\ref{kerrgen}), obtain finally the following solution to 
CEE (which is univalued everywhere on 3D-space):
\beg{secsol}
\Pi = -\frac{(r^*)^2}{w} = -\frac{x^2+y^2+(z+ia)^2}{x-iy}.
\enq
It is instructive to note that equation $\Pi=0$, being equivalent to two 
real-valued constraints $z=0,~x^2+y^2=a^2$, defines here the ring-like 
singularity for the Kerr solution (\ref{kerrsol}), as it should be in account 
of the theorem above presented (for this, see also section 4).

Static solutions of the II class with spatially bounded singularities 
are not at all exhausted by the solution (\ref{secsol}). Consider, for example,  
solutions generated by the functions
\beg{solint}
\Pi = \frac{G^n}{H}, ~~n\in \mathbb{Z},~~n>2 .
\enq
We'll not write out correspondent solutions in explicit form and shall restrict 
ourselves by examination of the spacial structure of their singularities which 
can be obtained from the joint system of equations $P=0,~\Lambda=0$, 
see Eqs.(\ref{caust}),(\ref{supsing}). Eliminating from the latter the unknown 
field $G$ we find that singularities (branching points of the eikonal field) 
have again the ring-like form $z=0,~x^2+y^2=R^2$ with radii equal to 
\beg{radii}
R_n = \frac{a (n-1)}{\sqrt{n(n-2)}}
\enq  
The cases $n=1,2$ evidently need special consideration. For $n=1$  
equating to zero derivative of the function $G/H$ we find $G=\pm i\bar w / \rho$ 
with $\rho =\sqrt{x^2+y^2}$. This brings us after substitution to the following 
solution of the CEE:
\beg{exepsol}
\Pi = (z+ia \pm i\sqrt{x^2+y^2})^{-1}
\enq
which has the pole at the ring $z=O, x^2+y^2 = a^2$ but {\it has a branching 
point only on the origin} $r=0$, i.e. {\it under any $a$ corresponds to the 
point singularity.}
                   
In the case $n=2$ via analogous procedure we get $G=\bar w / z^*$ and after 
substitution come to the following solution of the CEE~\cite{eik}:
\beg{hopf}
\Pi = \frac{\bar w}{r^*} = \frac{x+iy}{x^2+y^2+(z+ia)^2}
\enq
which is of the same structure as (the inverse of) the solution (\ref{secsol}).  
As the latter, it has no branching points on the real space-time slice while 
its pole corresponds to the Kerr ring. Let us take for simplicity $a=-1$; then 
solution (\ref{hopf}) can be rewritten in the following familiar form:
\beg{hopf1}
\Pi = i\frac{x+iy}{2z + i(r^2 - 1)}
\enq
which can be easily identified as the standard {\it Hopf map}. As the solution 
of the CEE it has been studied in the recent paper~\cite{adam} where its 
geometrical and topological nature has been 
examined in detail. We suspect also that generalized Hopf maps considered 
therein and obeying the CEE are closely related (in the case $m=1$) to the CEE 
solutions generated by the 
functions (\ref{solint}) and, as the latters, has the ring singularities 
correspondent to those represented by Eq.(\ref{radii}).
However, this should be verified by direct calculations. Note, finally, 
that trivial generalization of the functions (\ref{solint}) 
$\Pi=G^n/H^m, ~~n,m\in \mathbb{Z}$ gives rise to a set of CEE static solutions
with two integers which correspond, perhaps, to those introduced in~\cite{adam}.   

\noindent
{\bf Wave solutions.}  Consider also the class of generating functions dependent on 
one of the two twistor variables $\tau^0,\tau^1$ only, say on $\tau^0$:
\beg{wave}
\Pi = \Pi(G,\tau^0) = \Pi(G,u+wG).
\enq
Both classes of the CEE solutions obtained via functions (\ref{wave}) 
will then depend on only two spinor coordinates $u=t+z,~w=x-iy$. This means, in 
particular, that the fields propagate along the $Z$-axis with fundamental 
(light) velocity $c=1$. A ``photon-like'' solution of this 
type, with singular locus spacially bounded in all directions, was 
presented in~\cite{sing}.

Notice also that an example of the CEE solution with a considerably more rich 
and realistic structure of singular locus is presented below in section 4 
(see also~\cite{sing}). 

\section{Particles as caustics of the primodial light-like congruences}
 
It's well known that a null congruence of rays corresponds to any solution 
of the eikonal equation; it is orthogonal to hypersurfaces of constant 
eikonal $S = const$ and directed along the 4-gradient vector $\p_\mu S$. 
Usually, these two structures define the {\it characteristics} and 
{\it bicharacteristics} of a (linear) hyperbolic-type equation, e.g. of the 
wave equation $\square \Psi= 0$. 

In the considered complex case, i.e. in the case of CEE, the hypersurfaces of 
constant eikonal and the 4-gradient null congruences belong geometrically 
to the {\it complex extension} $\mathbb{C}M^4$ of the Minkowski space-time 
which looks here quite natural in account of the complex structure of the  
primary biquaternion algebra $\mathbb{B}$. 
The problem of physical sense of the additional (imaginary) dimensions is much 
important and nontrivial, and we hope to discuss it in the forthcoming paper. 

Here we use another interesting property: existence of a null geodesic 
congruence defined on a {\it real} space-time for every of the 
{\it complex-valued} solutions to CEE. This remarkable property follows 
directly from the twistor structure inherent to CEE. Indeed, according to 
the theorem above-presented, any of the CEE solutions (both of the I and the II 
classes) is fully determined by a (null projective) twistor field 
$\{\xi(X),\tau(X)\}$ (in the choosed gauge one has $\xi_0 = 1;~\xi_1 = G(x)$) 
subject to the incidence relation (\ref{inc}). This latter ``Penrose equation'' 
can be explicitly resolved with respect to the space coordinates 
$\{x_a,~a = 1,2,3\}$ 
as follows: 
\beg{coord}
x_a = \frac{\Im(\tau^+\sigma\xi)}{\xi^+\xi} - 
\frac{\xi^+\sigma\xi}{\xi^+\xi}~t~,
\enq
with $\{\sigma_a\}$ being the Pauli matrices and the time $t$ remaining a 
{\it free parameter}. Eq.(\ref{coord}) manifests that the primodial spinor 
field $\xi(X)$ {\it reproduces its value along the 3D rays formed by the unit 
``director vector'' 
\beg{dir}
\vec n = \frac{\xi^+\vec \sigma\xi}{\xi^+\xi}, ~~~ \vec n^2 = 1~,
\enq
and propagates along these locally defined directions with fundamental constant 
velocity} $c = 1$. In the choosed gauge we have for Cartesian components of the 
director vector (\ref{dir}) 
\beg{dircoord}
\vec n =\frac{1}{(1+GG^*)}\{(G+G^*),-i(G-G^*),(1-GG^*)\}, 
\enq
the two its real degrees of freedom being in one-to-one correspondence 
with the two components of the complex function $G(X)$. 

Thus, for every solution of the CEE 
the space is foliated by a congruence of {\it rectilinear} light rays, i.e 
by a {\it null geodesic~\footnote{On the flat Minkowsky background 
the geodesics are evidently rectilinear} congruence} (NGC). Notice that the 
director vector obeys the {\it geodesic equation}~\cite{number} 
\beg{geodes}
\p_t \vec n + (\vec n \vec \nabla)\vec n = 0 ~.
\enq

The basic field $G(X)$ of the NGC can be always extracted from one of the two 
algebraic constraints (\ref{kerr}) or (\ref{caust}) which at any space-time 
point possess, as a rule, 
not one but rather a {\it finite (or even infinite) set} of different solutions. 
Suppose that generating function $\Pi$ is {\it irreducible}, i.e. can't be 
factorized into a number of twistor functions of the same structure (otherwise, 
we should make a choice in favour of one of the multiplies). 
Then a generic solution of the constraints will be nothing but a {\it 
multivalued complex function} $G(X)$. 
Choose locally (in the vicinity of a particular point $X$) one of the 
continious {\it branches} of this function. Then a particular NGC and a set of 
physical fields can be associated with this branch, i.e. with  one of the 
``modes'' of the multivalued field distribution.  

Specifically, for any of the I class CEE solutions the spinor $F_{(AB)}$ of 
{\it electromagnetic field} can be defined explicitly in terms of twistor variables 
of the solution~\cite{jos,vest,sing}:
\beg{emsp}
F_{(AB)} = \frac{1}{P}\left\{\Pi_{AB} -\frac{d}{dG}\left(\frac{\Pi_A\Pi_B}{P}
\right)\right\} .
\enq
where $\Pi_A,\Pi_{AB}$ are the first and the second order derivatives of the 
generating function $\Pi$ with respect to its two twistor arguments $\tau^0, 
\tau^1$. For every branch of the solution $G(X)$ {\it this field locally satisfies
Maxwell homogeneous (``vacuum'') equations.} Moreover, as it has been demonstrated 
in~\cite{GR,kasan,jos}, a complex-valued $SL(2,\mathbb{C})$ {\it Yang-Mills 
field} and a {\it curvature field} (of some effective Riemannian metric)
can be also defined through only the same principal function $G(X)$ for any 
of the CEE solution of the first class.

Consider now analytical continuation of the function $G(X)$ up to one of its 
branching points which corresponds to a multiple root of Eq.(\ref{kerr}) (or, 
alternatively, of Eq.(\ref{caust}) for solutions of the II class). At this point
$P=0$, and the strength of electromagnetic field (\ref{emsp}) turns 
to infinity. The same holds for the other associated fields, for curvature field
~\footnote{Associated Yang-Mills fields possess, generically, 
additional {\it string-like} singularities} in particular~\cite{burin}. Thus, 
the locus of branching points (which can be 0-, 1- or even 2-dimensional, see 
section 4) manifests itself as a {\it common source} of 
a number of physical fields and can be identified (at least, in the case when 
it is bounded in 3-space) as a unique {\it particle-like} object.

Such formations are capable of much nontrivial evolution simulating physical 
interactions or even mutual {\it transmutations} represented by {\it bifurcations} 
of the field singularities (see, e.g., the example in section 4). They possess 
also a realistic set of ``quantum numbers'' including a {\it self-quantized 
electric charge} and a {\it Dirac-type gyromagnetic ratio} (equal to 
that for a spin 1/2 fermion)~\cite{carter,lopes,prot}. Numerous examples of 
such solutions and their singularities can be found in our works~\cite{kasan,
prot,trish,jos}.

On the other hand, for the light-like congruences - NGC - associated with CEE 
solutions via the guiding vector (\ref{dircoord}) the locus of branching 
points coincides with that of the principal $G$-field and represents  
the familiar {\it caustic} structure, i.e. the envelope of the system of rays 
at which the neighbouring rays intersect each other (``focusize''). From this viewpoint, 
within the algebrodynamical theory {\it the ``particles'' are nothing but the 
caustics of null rectilinear congruences associated with the CEE solutions.}

\section{The World function and the multivalued physical fields}

At this point we have to decide which of the two types of the CEE solutions 
can be in principle taken in our scheme as a representative for description of 
the Universe structure as a whole. {\it As a ``World solution'' we choose a CEE 
solution of the first class} because a lot of peculiar geometrical structures
and physical fields can be associated with any of them~\cite{GR,prot,jos}. 
Such a solution can be obtained algebraically from the Kerr functional 
constraint (\ref{kerr}) and a generating twistor ``World function'' $\Pi$ 
which is {\it exceptional} with respect to its internal properties; 
geometrically it gives rise to an NGC with a special property - zero shear~\cite
{penrose,raygeom}.

Moreover, a {\it conjugated} CEE solution of the II class turns then also to be 
involved into play since it defines a characteristic hypersurface of the 
(I class) ``World solution''. In fact, this is determined
as a solution of the joint algebraic system of Eqs.(\ref{kerr}),(\ref{caust}). 
Precisely, if we resolve Eq.(\ref{caust}) with respect to $G$ and substitute the 
result into (\ref{kerr}), equation $\Pi(G(X)) = 0$ would define then the 
singular locus (the characteristic hypersurface) of the World solution. 
On the other hand, the function $\Pi(G(X))$ would necessarily satisfy the CEE 
representing its II class solution in account of the theorem presented
in section 2. Thus,

\noindent
{\it the eikonal field here carries out two different functions being a 
fundamental physical field (as a CEE solution of the I class) and, at the same 
time, a characteristic field (as a solution of the II class) which
describes the locus of branching points of the basic field (i.e., the 
discontinuties of its derivatives).}

Let us conjecture now that the World function $\Pi$ is {\it an irreducible 
polinomial of a very high but finite order}~\footnote{This conjecture is,  
in fact, not at all necessary. Indeed, one can easily imagine that the World  
function leads to the Kerr Eq.(\ref{kerr}) which possesses an {\bf infinite 
number of roots} for complex-valued field function $G(X)$ at any space-time point $X$}
so that Eq.(\ref{kerr}) is an algebraic (not a transcendental) one. 
Note that in this case Eq.(\ref{kerr}) defines 
an algebraic surface in the projective twistor space $\mathbb{C}P^3$.

The World solution consists then of a finite number of modes -- 
branches of multivalued complex $G$-field. A finite number of null directions 
(represented in 3-space by the director vector (\ref{dircoord})) and an equal 
number of locally distinct NGC would exist then {\it at every point}.

Any pair of these congruences at some fixed moment of time will, generically, 
has an envelope consisting of a number of connected one-dimensional 
components-caustics~\footnote{In fact, the caustics of {\bf generic type} 
are determined by one complex condition $\Pi(G(X)) = 0$ (i.e., by two real
equations) on three coordinates and, at a fixed moment of time $t = t_0$, 
correspond to a number of one-dimensional curves (``strings'')}.
Just these spacial structures (in the case they are bounded in 3-space) 
represent here the ``particles'' of generic type. Other types
of particle-like structures are formed at the focal points of {\it three or 
more} NGC where Eq.(\ref{kerr}) has a root of higher multiplicity. Formations 
of the latter type would, of course, meet rather rarely, and their
stability is problematic. One can speculate on their possible relation to 
particle's excitations -- {\it resonances}.

Nonetheless, we can model both types of particles-caustics in a simple example 
based on generating twistor function of the form~\cite{sing}
\beg{cocoon}
\Pi = G^2(\tau^0)^2+(\tau^1)^2 - b^2G^2 = 0, ~~~b=const\in \mathbb{R}~,
\enq
which leads to the 4-th order polinomial equation for the $G$-field. At initial 
moment of time $t = 0$, as it can be obtained analytically, the singular locus 
consists of a pair of point singularities (with opposite and
equal in modulus ``elementary'' electric charges) and of a neutral 2-surface 
(ellipsoidal {\it cocoon}) covering the charges (see~\cite{sing} for more 
details). The latter corresponds to the intersection of all of the 4 modes
of the multivalued solution while each of the point charges is formed by 
intersection of a particular pair of (locally radial, Coulomb-like) congruences
~\cite{sing}. The time evolution of the solution and of its singularities is 
very peculiar: for instance, at $t = b/\sqrt{2}$ the point singularities cancel 
themselves at the origin $r = 0$ simulating thus the process of {\it 
annihilation} of elementary particles. Moreover, this process is accompanied 
by emission of the {\it singular light-like wavefront} represented by another 
2-dimensional component of connection of the caustic structure.

Thus, we see that the multivalued fields are quite necessary for to ensure 
the self-consistent structure and evolution of a complicated (realistic) system 
of particles - singularities. One only should not be confused by such, much 
unusual, property of the principal $G$-field and, especially, by multivalued 
nature of the other associated fields including the electromagnetic one.

Indeed, in convinient classical theories, the fields are in fact only a tool 
which serves for adequate description of particle dynamics (including the 
account of retardation etc.) and for nothing else. In nonlinear theories, 
as well as in our algebrodynamical scheme, the fields are moreover 
responsible for {\it creation} and structure of particles themselves, as 
regular {\it solitons} or {\it singularities} of fields respectively. In the 
first, more familiar case we, apparently, should consider the fields to be 
univalued. The same situation occures in the framework
of quantum mechanics where the quantization rules often follow from the 
requirement for the wave function to be univalued.

However, as we have seen above, in the algebrodynamical construction {\it the 
field distributions must not necessarily be univalued !} On the other hand, 
acception of fields' multivaluedness does not at all prevent to obtain the 
discrete spectrum of characteristics in a full analogy with quantum mechanics. 
For example, the requirement of univaluedness of a {\bf particular, locally 
choosed mode} of the principal $G$-field and of the associated electromagnetic 
field (far from the branching points of the first and, consequently, from the 
infinities of the second!) leads to the general property of {\it quantization of 
electric charge of singularities} in the framework of algebrodynamical 
theory~\cite{vest,sing}.

As to the process of ``measurement'' of the field strength, say, of 
electromagnetic field, it directly relates to only the measurements of particles' 
accelerations, currents etc., and only after the measurements the results are 
translated into conventional field language. However, this is not at all 
necessary (in recall, e.g, of the Wheeler-Feynman electrodynamics and of 
numerous ``action-at-a-distance'' approaches~\cite{smirnov,vladimirov}). 
In fact, ``we never deal with fields but only with particles'' (F. Dyson).

In particular, on the classical (nonstochastic) level we can deal, effectively, 
with the  {\it mean value} of the set of field modes at a point; similar concept 
based on purely quantum considerations has been recently developed in the 
works~\cite{kirill}. In our scheme, the true role of the multivalued field 
will become clear only after the spectrum and the effective mechanics of 
particles-singularities will be obtained in a general and explicit form.

We hope that a sort of psycological barrier for acception of general 
idea of the field multivaluedness will be get over as it was with possible 
{\it multidimensionality} of physical space-time. The advocated concept seems 
indeed very natural and attractive. In the purely mathematical framework,
multivalued solutions of PDEs are the most common in comparison with the 
familiar $\delta$-type distributions~\cite{lychagin,vinogr}. From 
physical viewpoint, this makes it possible to naturally define a dualistic
``corpuscular-field'' complex of a very rich structure which, actually, gathers 
all the particles in the Universe into a unique object. The caustics-singularities 
are well-defined themselves and undergo a collective self-consistent motion 
free of any ambiguity or divergence (the latters can arise here only in result of  
incorrect discription of the evolution process and can be removed, if arise, 
on quite legal grounds, contrary, say, to the renormalization procedure in 
the quantum field theory). Note also that recently accomplished universal 
local classification of singularities of differentiable maps, in particular 
of caustics and wavefronts~\cite{arnold2}, can explicitly bear on the 
characteristics of elementary particles if the latters are treated in the 
framework of the algebrodynamical theory. 

As to the principal problem of the choice of a particular representative of 
the generating {\it World function $\Pi$ of the Universe} we are 
ready to offer an interesting candidature being in hope to discuss it 
elsewhere.

\section{The light-formed relativistic aether and the nature of time}

Light-like congruences (NGC) are the basic elements of the picture of physical 
world which arises in the algebrodynamical scheme and, to some extent, in 
twistor theory in general. The rays of the NGC densely fill the space and
consist of a great number of branches - components superposed at each space 
point and propagating in different directions with constant in modulus and 
universal (for any branch of multivalued solution, any point and any system of 
reference) fundamental velocity. {\it There is nothing in the Universe exept
this primodial light flow (``pre-Light Flow'')} because {\it the whole Matter 
is born by pre-Light and from pre-Light} at the caustic regions of 
``condensation'' of the pre-light rays.

In a sense, one can speak here about an exeptional form of {\it relativistic 
aether} which is formed by a flow of pre-Light. Such an exeptional form of the 
World aether has nothing in common with old models of the {\it light-carrying} 
aether which had been considered as a sort of elastic medium. Here, the {\it 
light-formed} aether consists of structureless ``light elements'' and is, 
obviously, in full correspondence with special theory of relativity~\footnote
{At present, it seems rather strange that A. Einstein didn't come himself to 
the concept of relativistic aether so consonant with the ideas of STR and 
with his favourite {\it Mach principle}. Suprisingly, R. Penrose also 
overlooked this opportunity which follows naturally from his twistor theory}.  

At the same time, notions of the aether formed by pre-Light and of the matter 
formed by its ``thickenings'' 
evoke numerous associations with the Bible and with ancient Eastern philosophy. 
Certainly, there were teologists, philosophers or mystics who were brought to 
imagine a similar picture of the World. However, in the framework of successive 
physical theory this picture becomes more trushworthy and, to our knowledge, 
has not been yet discussed in literature~\footnote{Similar in some aspects 
ideas have been advocated in the works~\cite{urusovski,smolyanin,shelaev}. 
Note, in particular, the concept of the ``radiant particle'' 
offered by L.S. Shikhobalov~\cite{shihobal}}.

On the other hand, existence of the primodial light-formed aether and 
manifestation of universal property of local ``transfer'' of the {\it aether - 
generating field} $G(X$) with constant fundamental velocity $c=1$ points to 
different status of space and time coordinates and {\it offers a new approach 
to the problem of physical time} as a whole. By this, it is noteworthy that 
since in 1908 H. Minkowsky has joined space and time into a unique 
4-dimensional continuum, no further understanding of the nature of time 
has been achieved in fact. Moreover, this synthesis has ``shaded'' the principal 
distinction of space and time entities and clarified none of such 
problems as (micro/macro)irreversibility, (in)homogeneity and (non)locality 
of time, its dependence on material processes etc.

In the interim, the key problem of Time can be formulated in a rather simple way. 
{\it Subjectively}, we perceive time as a continious intrinsic motion, a
latent {\it flow}. Everybody comprehends in a moment, as the ancient Greeks did, 
what is meant by the ``River of Time'', the ``Flow of Time''. As a rule, we 
consider this intrinsic motion to be independent on our will and on material 
processes and uniform: not for nothing, in physics the flow of time is modelled 
by the uniform motion of, say, the record tape etc. Moreover, 
under variations in time one does not only observe the {\it conservation} of 
a particular set of {\it integral} quantities  
(which is widely used in the orthodox physics) but perceives subjectively the 
complete {\it repetition, reproduction of the {\it\bf local states} of any 
system;} that's why for measurements of time itself we use {\it clocks} whose 
principle of operation is based on reproducible, periodical processes. In other 
words, whereas one has much ambiguious and diverse distributions of spacial 
positions of physical bodies, all they and we all have always {\it one and the 
same monotonically increasing time coordinate}, i.e. are in a common and 
permanent motion together with the ``Time River''. 

Surprisingly, almost all these considerations are absent in the structure of 
theoretical physics and, in particular, in relativity theory. To bring 
into correspondence the results of calculations with practice (e.g. for the 
Cauchy problem etc.) one chooses a ``time orthogonal hypersurface'', i.e. 
quite ambiguiously fixes the unity of the present moment of time, of the moment  
``now'',  perceived subjectively by everybody; however, {\it there are no 
intrinsic reasons for this choice in the very structure of theoretical 
physics , including the STR.}
                             
At least partially, such a situation is caused by the following. The notion of 
everywhere existing, eternal Flow of Time immegiately leads to the problem of its 
(material? pre-material?) carrier. In this connection, the works 
of N.A. Kozy'rev~\cite{kozyr} should be marked, of course, in which the concept 
of the ``active'' Flow of Time influencing directly the material processes has been 
proposed. To our opinion, however, there are no reliable physical grounds 
at present which confirm the Kozy'rev's ideas, and no mechanism of ``interaction'' 
of this exotic form of matter with the ordinary ones. As to the 
algebrodynamical paradigm, the Time Flow is non-material therein: it 
does not {\it interact} or {\it influence} the Matter at all but just {\it forms} 
it. In distinction from the Kozy'rev's concept, we do not deal here with various  
material entities only one of them being the Time itself: on the contrary, here we 
have one triply-unique entity -- preLight-Time-Matter~\footnote{ 
Much more close the approach turns to be to the concept of 
``Time-generating Flows'' developed by A.P. Levich~\cite{levich}.  Projective structure 
of a specific type closely related to twistors has been applied for explanation 
of the time nature in the concept of M. Saniga~\cite{saniga}}.

On the other hand, under consideration of the problem of the carrier of the Time 
Flow, we inevitably return back  to the notion of some form of the {\it World 
aether} which has been exiled from physics after the triumph of Einstein's theory. 
To do without aether, none Flow of Time can be successively included into 
the structure of theoretical physics and  none subjectively perceived properties 
of time can be precisely formulated and described.

However, in a paradoxical way, just the STR with its postulate of the 
invariance of light velocity justifies the introduction of the 
{\it dynamical Lorentz invariant aether} formed by the light-like congruences 
as the primary element of physical World. Specifically, the Time Flow 
can be naturally identified now with the Flow of Primodial Light (pre-Light), 
and the ``River of Time'' turns to be nothing but the ``River of Light''. 
Moreover, it is the universality of light velocity which explains our 
subjective perception of uniformity and homogeneity of the Time Flow. 
  
There is, however, another, the most striking and unexpected feature of the 
introduced concept of physical time. {\it The Time Flow manifests here itself 
as a {\bf\it superposition} of a great number of distinctly directed and 
locally independent components - ``subflows''}. At any point of 3-dimensional 
space there exists a (finite) set of directions: each mode of the 
primodial multivalued field $G(X)$ defines one of these directions and 
{\it propagates (reproduces its value) along it} forming thus one of the 
constituents of the (globally unique) Flow of pre-Light identical to the Flow 
of Time.   

One can conjecture that just by virtue of the local multivaluedness we are 
not capable of to perceive the particular local {\it direction} of the Time 
Flow. Apart from this, it is natural to assume that in the tremendously 
complicated structure of the World solution a {\it stochastic} component is 
necessarily present, particularly in the structure of the primodial Light-Time 
Flow. This results in chaotic variations of local directions of the 
light-like congruences which are certainly inaccessible for perception. On the 
other hand, it is the existence of (constant in modulus and the same for all 
of the branches of the multivalued World solution) {\it fundamental propagation 
velocity of the pre-Light rays} which makes it possible to feel the Flow of Time 
in general and to subjectively regard it as uniform and homogeneous in particular.

\section{Conclusion}

Thus, we have examined the realization of the algebrodynamical approach 
in which as a base of unified physical theory one only structure of a purely 
abstract nature is choosed, namely the algebra of complex quaternions and the 
generalized CR-equations -- the conditions of differentiability in this algebra.
Very the same structure can be successively expressed, in fact, on a number 
of equivalent geometrical languages (covariantly constant fields, twistor 
geometry, shear-free congruences etc.). 

Primary GCR-equations result directly in the field of complex eikonal
regarded in theory as a fundamental physical field (alternative in a sense to 
the linear fields of quantum mechanics). In its turn, the eikonal field 
is here closely related to the fundamental 2-spinor and twistor fields, on 
whose language, in particular, the general solution of the complex eikonal  
equation is formulated. Through the eikonal field also the other ones
are defined, namely the electromagnetic and Yang-Mills fields. Singularities 
of the eikonal and of correspondent null congruences are considered as 
particle-like formations (``self-quantized'' and effectively interacting).  

In result, physical picture of the World which arises as a consequence of 
one only algebraic structure appears as very beatiful and unexpected. As 
its basic elements it contains the primodial light flow -- ``pre-Light'' -- 
and the relativistic aether formed by the latter, multivalued physical fields and 
prelight-born matter (consisting of particles-caustics formed by 
the superposition of individual branches of the unique pre-light congruence 
in the points of their ``focusization'').

As very natural and deep seems to be the arising in theory connection between 
the existence of universal velocity (velocity of ``light'') and of the time 
flow; connection which permits to understand, in a sense, the origin of the 
Time itself.  {\it Time is nothing but the primodial Light}; these two entities 
are undividible. On the other hand, {\it there is nothing in the World except 
the preLight Flow} which gives rise to all the ``dense'' Matter in the Universe.
\section{Acknowledgements}

The author is grateful to D.P. Pavlov for his invitation to contribute into the 
new and actual journal ``Hypercomplex numbers in physics and geometry'', and 
to participate in competition of scientific works on related topics. I am also 
greatly indebted to A.P. Levich and to the participants of scientific seminar 
on the study of time phenomenon supervised by him. Much fruitful were  
my conversations with V.I. Zharikov, V.I. Zhuravlev, J.A. Rizcallah, V.N. Trishin, 
V.P. Troitsky, V.P. Tsarev and other my colleagues to whom I am deeply 
grateful for their long-termed friendship and support. I indeed beleive that 
tremendous building of conteprorary physics can be completely reconstructed 
to a ``new design'' which is much simple, the ``only possible'' (J.A. Wheeler) 
and which brings us nearer to the true Project in accord to which our World 
has been created once upon a time.

\end{document}